\documentclass[review,12pt]{elsarticle}
\usepackage[margin=2.5cm]{geometry}
\usepackage{lineno,hyperref}
\usepackage{soul}
\modulolinenumbers[5]

\usepackage{xcolor}

\usepackage{rotating}
\usepackage[notablist,nofiglist,nomarkers]{endfloat}
\DeclareDelayedFloatFlavor{sidewaystable}{table}

\usepackage{amsmath}
\usepackage{amssymb}
\usepackage{wasysym}
\usepackage{rotating}

\biboptions{sort&compress}
\makeatletter
\def\ps@pprintTitle{%
 \let\@oddhead\@empty
 \let\@evenhead\@empty
 \def\@oddfoot{}%
 \let\@evenfoot\@oddfoot}
\makeatother

\begin{document}

\begin{frontmatter}

\title{Nucleation properties of isolated shear bands}


\author{Shwetabh Yadav, Dinakar Sagapuram$^1$}

\address{Department of Industrial and Systems Engineering, Texas A\&M University, College Station, TX, USA}

\fntext[myfootnote]{Corresponding author, E-mail: dinakar@tamu.edu}






\begin{abstract}
\noindent
Shear banding, or localization of intense strains along narrow bands, is a plastic instability in solids with important implications for material failure in a wide range of materials and across length-scales. In this paper, we report on a series of experiments on the nucleation of single isolated shear bands in three model alloys. Nucleation kinetics of isolated bands and characteristic stresses are studied using high-speed \emph{in situ} imaging and parallel force measurements. The results demonstrate the existence of a critical shear stress required for band nucleation. The nucleation stress bears little dependence on the normal stress and is proportional to the shear modulus. These properties are quite akin to those governing the onset of dislocation slip in crystalline solids. A change in the flow mode from shear banding to homogeneous plastic flow occurs at stress levels below the nucleation stress. Phase diagrams delineating the strain, strain rate and temperature domains where these two contrasting flow modes occur are presented. Our work enables interpretation of shear band nucleation as a crystal lattice instability due to (stress-assisted) breakdown of dislocation barriers, with quantitative experimental support in terms of stresses and the activation energy.

\end{abstract}


\end{frontmatter}

\renewcommand{\efloatseparator}{\vfill{}}

\newpage

\section{Introduction}

When ductile metals are subjected to large plastic strains, particularly at high strain rates ($\gtrsim 10^3$ /s), it is often found that the deformation pattern is not smooth and homogeneous, but shows marked localization of strain along narrow bands \cite{dodd2012}. This phenomenon, called shear banding, has important implications since  strain localization often acts as a prelude to material failure. This understandably has adverse repercussions for material's ductility and integrity in deformation processing \cite{hatherly1984, lapovok2009, azizi2018} and high-rate structural applications such as penetration and impact \cite{timothy1985, wright2002, rittel2008}. Clearly, a detailed understanding of shear band onset characteristics is of critical importance both for quantifying conditions required for banding and designing practical means to control it \cite{rice1976, burns1997, wright2002}. While most extensively studied in the context of polycrystalline metals, shear banding has been also widely observed in other material systems such as metallic glasses \cite{ donovan1981, lewandowski2006, jiang2009, jiang2011}, polymers \cite{bowden1970, fleck1990} and granular media \cite{palmer1973, walley2012}.

The actual appearance of shear bands are dependent to some extent on the material itself, and comprehensive reviews of shear bands can be found in Refs.~\cite{dodd2012, antolovich2014, walley2007}. Nevertheless, recent \emph{in situ} observations \cite{sagapuram2016, yadav2020} suggest that the sequence of steps leading to a fully developed band with high strain localization in its vicinity can be demarcated into distinct stages of \emph{nucleation} and \emph{growth}, irrespective of the material system. During the nucleation stage, a strain inhomogeneity nucleates in an otherwise homogeneous strain field, followed by rapid propagation of its front to establish a thin and well-defined shear band interface across the whole specimen. The subsequent growth stage is characterized by accumulation of large plastic strains in the immediate vicinity of this  interface, while regions away from the band essentially slide as rigid bodies. In this picture, the minimum shear stress required to nucleate a thin interface may be regarded as the nucleation stress, and stress that drives the localized flow development as the growth stress, of shear bands themselves \cite{sagapuram2018}.

The present paper is concerned with understanding the nucleation properties of shear bands in ductile metals, with a specific focus on nucleation stresses. Whereas considerable progress has been  made in modeling shear band nucleation and propagation aspects \cite{cherukuri1997, grady1992,bigoni2008, bordignon2015}, direct experimental characterization of band nucleation  has proved a difficult problem. This is primarily because  high strain-rate loading configurations such as impact, dynamic compression-shear testing, punch testing and explosive loading of thick-walled cylinders typically used to study shear banding are not suited for isolation of single bands. Furthermore, the small timescales (typically a few $\mu$s) involved  in band nucleation pose additional experimental challenges. As a result, measurements of band nucleation properties (dynamics, stresses, etc.) in metals have been quite limited. It may be also noted that, even in metallic glasses where shear band nucleation stresses are often reported, stresses are not directly measured but usually extracted from the nanoindentation tests \cite{packard2007, perepezko2014, wang2011}.

In our prior work, we have devised a novel 2D shear configuration wherein isolated shear bands can be formed and examined independently without interfering effects from other bands \cite{sagapuram2016, sagapuram2018}. This has enabled \emph{in situ} observations of single shear bands, and elucidated several key features of shear banding such as band nucleation, propagation kinetics and kinematics of localized flow (boundary layer) evolution during the secondary growth stage \cite{yadav2020}.

In the present study, using this configuration, we probe nucleation properties of isolated shear bands in three different model, low melting-point alloys. \emph{In situ} high-speed imaging and force measurements are used to decouple shear band nucleation from the subsequent growth stage, and thereby accurately determine  equivalent stresses. The principal finding from our study is that the band nucleation stress $\tau_C$ is a material-dependent constant that is insensitive to external loading conditions and  $\approx  0.05 \mu_0 $, with $\mu_0$ being the shear modulus. A transition in the flow pattern from shear banding to homogeneous-type at stress levels below $\tau_C$ is also unambiguously demonstrated using our \emph{in situ} flow observations. Based on the results, phase diagrams demarcating shear banding and homogeneous flow modes are constructed.

The remainder of the paper is organized as follows. The shear loading setup, imaging methods and material systems used in the study are described in Sec.~\ref{sec:experimental}. \emph{In situ} observations of shear band dynamics and calculation of nucleation stresses are presented in Sec.~\ref{sec:results}. The transition between shear banding and homogeneous flow modes, and phase diagrams showing where each of these regimes occur, are also presented in this section. Implications of the findings and a microscopic interpretation of shear band nucleation as a mechanical instability involving sudden breakdown of dislocation barriers are discussed in Sec.~\ref{sec:discussion}. Concluding remarks are presented in Sec.~\ref{sec:conclusions}.

\section{Experimental} \label{sec:experimental}
\begin{figure}
\centering \includegraphics[width=4.0in]{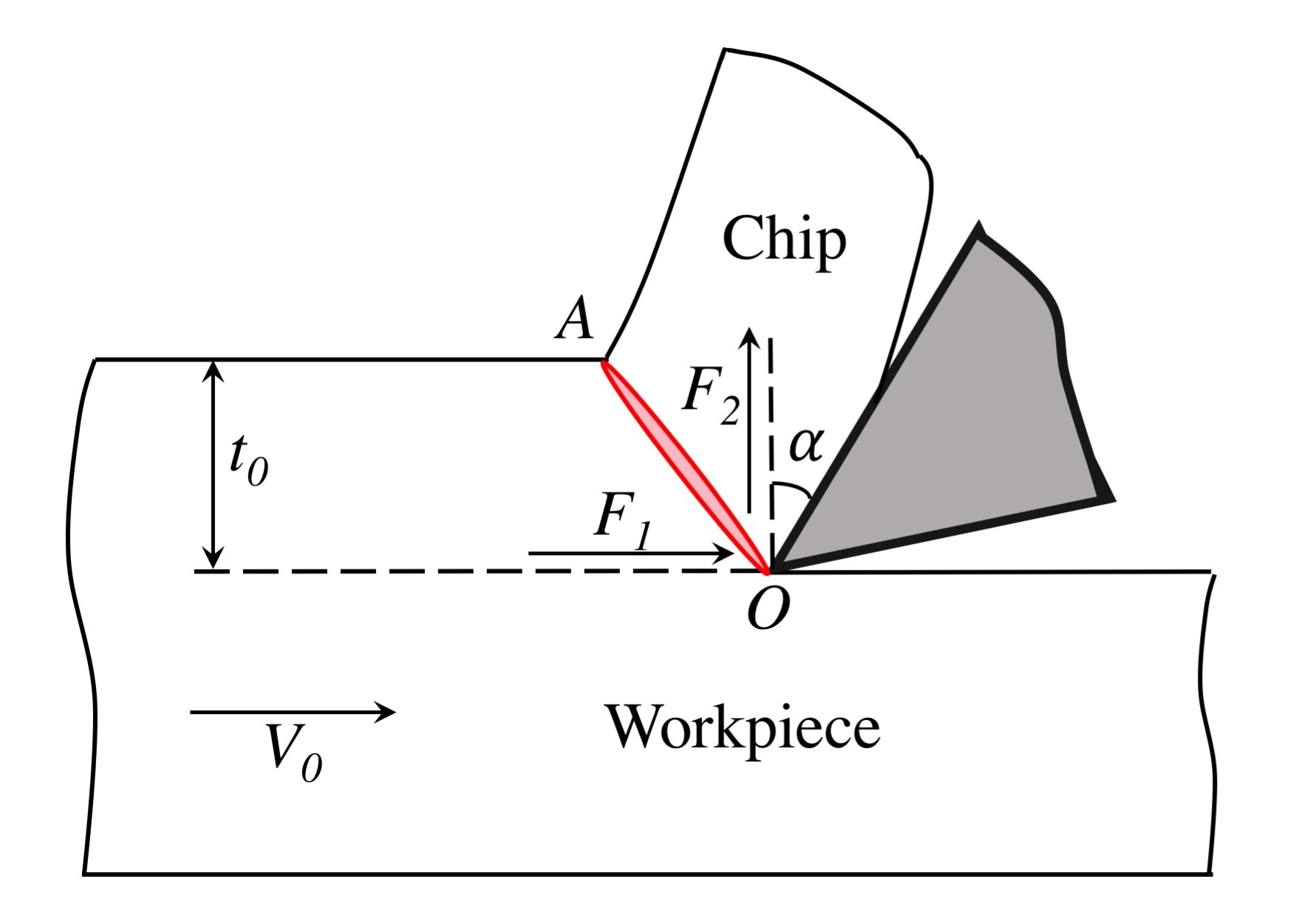}
\caption{\label{fig:cutting} Schematic of  2D cutting configuration used to impose large-strain shear deformation. The deformation zone ($OA$) where plastic shearing occurs, as the material is transformed into a chip, is highlighted in red. Two orthogonal force components, $F_1$ and $F_2$, acting on the tool are shown, along with the experimental parameters ($\alpha$, $V_0$ and $t_0$).}
\end{figure}

The loading configuration used to study shear bands is a 2D (plane-strain) cutting configuration, as shown in Fig.~\ref{fig:cutting}. In this configuration, material is subjected to nominal simple shear in a narrow deformation zone (red shaded area \emph{OA}, Fig.~\ref{fig:cutting}) as a thin layer of material of predefined thickness $t_0$ is removed in the form of a chip by a sharp wedge-shaped tool. The tool inclination $\alpha$ (Fig.~\ref{fig:cutting}) controls the level of imposed shear strain ($\gamma$), while strain rate  ($\dot \gamma$) in the deformation zone is primarily determined by the deformation velocity $V_0$. For instance, shear strains in the range of 1-5 and strain rates from 10 to $10^5$ /s can be conveniently accessed by controlling these parameters. More importantly, for the present purposes, this shear configuration allows formation of isolated shear bands, i.e., one at a time, and at an \emph{a priori} known location (tool tip) \cite{sagapuram2018, yadav2020}. This feature is central to the present study as it allows shear band nucleation to be directly examined \emph{in situ}. Furthermore, a large number of shear bands ($> 200$) can be generated in a single experiment or ``cut", which offers additional benefits in regards to data repeatability verification and statistical analysis of band nucleation attributes such as stresses and propagation velocity. Some of these unique aspects of the cutting geometry in relation to study of shear bands appear to have been recognized more than 50 years ago  by Recht \cite{recht1964}. Several recent studies, e.g., Refs.~\cite{yeung2015,guo2015,zeng2018}, also suggest a renewed interest in utilizing cutting-deformation framework for studying  large-strain deformation phenomena, including shear banding, flow instability and ductile fracture. Similar 2D configurations have also proved invaluable for isolating frictional  wave phenomena at soft adhesive interfaces \cite{viswanathan2016}.

The material systems studied are three low melting-point ($T_m$) bismuth-based eutectic alloys (labeled Alloys 1-3) with melting points of 47 $^\circ$C, 70 $^\circ$C and 138 $^\circ$C, respectively. The nominal composition and thermophysical properties of these alloys are given in Table~\ref{tb:properties}. The choice of these alloys was governed by the fact that they exhibit shear banding  at strain rates multiple orders smaller than the typical rates ($> 10^3$ /s) at which banding occurs in engineering alloys such as Ti and steels \cite{yadav2020}. This means that shear band dynamics can be studied \emph{in situ} at high resolution, both spatially and temporally, which is highly difficult, if not impossible, to achieve in conventional high $T_m$ materials. Furthermore, the deformation behavior of these alloys under ambient temperature conditions is such that the flow stress is highly sensitive to strain rate, with little dependence on strain; for example, see compression test data provided in Fig.~S1 (Supplementary Material). In all three alloys, the ambient temperature flow stress $\sigma$ varies with effective strain rate ($\dot\varepsilon$) as: $\sigma \propto \dot\varepsilon^m$, where $m$ is in the range of 0.06-0.13 (depending on the alloy). It is important to note that this type of highly rate-dependent behavior (with negligible strain hardening) is a common characteristic of metals deforming at very high strain rates, typically $> 10^3$ /s \cite{zener1944}.  Therefore, low $T_m$ alloys are attractive model materials to experimentally `simulate' the most essential features of high strain-rate behavior of metals (including shear banding) at low deformation speeds. 
 
The materials were obtained from Belmont Metals Inc. (Brooklyn, NY) in the form of ingot and cast into desired shape and size by preheating the metal to 200 $^\circ$C and pouring into a rectangular-shaped aluminum molds ($\sim 75$ mm $\times$ 25 mm $\times$ 2 mm). Wood's metal (Alloy 2, $T_m = 70$ $^\circ$C) was chosen as the prototypical material in most of the experiments. The shearing experiments were carried out using a freshly ground high-speed steel tool (edge radius $\sim 5$ $\mu$m) at different $V_0$ in the range of  0.005-10 mm/s. This corresponds to a strain rate variation  over 4 orders of magnitude from $\sim 0.1$ to 10$^3$ /s. The $t_0$ was kept in the range of 150 $\mu$m to 500 $\mu$m, and $\alpha$ between $-20^\circ$ to $+40^\circ$. The effects of ambient temperature were studied by performing experiments at different pre-heat/cool temperatures ($T_0$) from $-20$ $^{\circ}$C to $65$ $^{\circ}$C. In these experiments, both the sample and tool-holder assembly were maintained at the desired temperature using dry ice or a heat gun. 

In all the experiments, force components parallel and perpendicular to $V_0$ direction (Fig.~\ref{fig:cutting}) were measured using a piezoelectric force sensor (Kistler 9129AA) mounted directly under the tool. These measurements were used for stress calculations. 

Direct time-resolved observations of the plastic flow were made using a high-speed CMOS camera (pco dimax HS4), synchronized with the force sensor. The plane-strain condition at the specimen side-surface being imaged was ensured by lightly constraining this side using a transparent sapphire plate and imaging through this plate \cite{yadav2020}. The spatial  resolution of our imaging was 0.98 $\mu$m (per pixel). Although the camera is capable of recording image sequences at up to 50,000 frames per second,  frame rates with a 100-200 $\mu$s inter-frame time (depending on $V_0$) were found to be adequate for capturing the shear band dynamics, especially given the low deformation speeds under which shear bands can be produced in low $T_m$ alloys \cite{yadav2020}. Quantitative full-field displacement data were obtained by analyzing the high-speed image sequence using  a correlation-based image processing method called as Particle Image Velocimetry (PIV) \cite{adrian2010}. Using PIV displacement data, the plastic flow field characteristics were analyzed using streaklines, grid deformation, strain and strain rate maps. This imaging and image analysis also enabled us to establish direct correlations between the macroscopic force traces and shear band dynamics at the mesoscale. A detailed description of our experimental setup and full-field deformation analysis can be found in Ref.~\cite{yadav2020}.

\section{Results}\label{sec:results}

 \emph{In situ} imaging, image analysis and force measurements have provided complete characterization of the shear band dynamics, and enabled characterization of nucleation stresses across different strain rates and temperatures. 
\subsection{General attributes of shear banding} \label{sec:general_features}
\begin{figure}
\centering \includegraphics[width=6.5in]{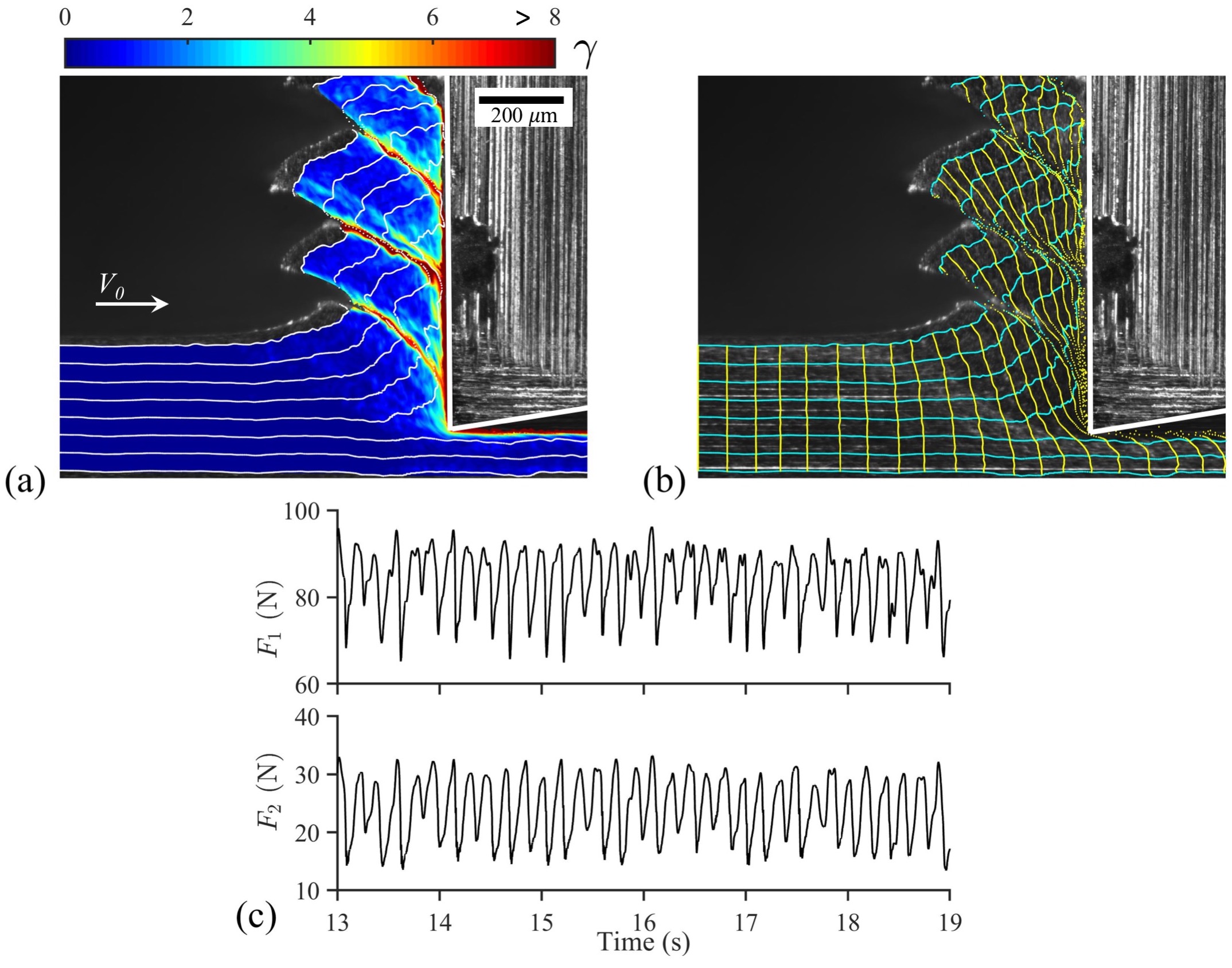}
\caption{\label{fig:shearbands}Shear banding characteristics at $V_0$ = 1.5 mm/s:  (a) shear strain field and (b) artificial grid superimposed on the high-speed image showing a highly heterogeneous plastic flow characterized by high-strain shear bands and  surrounding low-strain regions. The corresponding force vs. time plot is shown in (c), where each oscillation in the force trace reperesents single shear band formation.}
\end{figure}

We begin with a synopsis of the primary experimental observations pertaining to shear band attributes and their dynamics. Figure~\ref{fig:shearbands} shows a snapshot of a typical experiment ($\alpha$ = 0$^\circ$, $t_0$ = 250 $\mu$m) where the flow is dominated by periodic shear banding. In this experiment, $V_0$ = 1.5 mm/s, which corresponds to a nominal strain rate of $\sim 50$ /s. Figure~\ref{fig:shearbands}(a) shows the PIV-measured shear strain field and streaklines (white lines) superimposed on a high-speed image. The deformed material (chip) is clearly characterized by two distinct regions --- thin bands of high strain separated by low-strain segments. The typical shear band strain is $\sim 8$, an order of magnitude higher than in the surrounding regions. The large strain localization within shear bands can be also seen from Fig.~\ref{fig:shearbands}(b), which shows an artificial deformed grid on top of the image. This was obtained by initially overlapping a square grid (40 $\mu$m $\times$ 40 $\mu$m) on the undeformed workpiece and tracking/updating each point on the grid lines during the deformation process. The stark contrast in the extent of deformation between shear bands and surrounding regions is evident. While the band material is highly sheared, as can be judged based on relative positions of streaklines (cyan color, Fig.~\ref{fig:shearbands}(b)) on either side of the band, material segments in between the bands are only slightly distorted. It should be also noted that shear bands seen here are macroscopic, in the sense that they traverse across the entire specimen width (dimension normal to the viewing plane in Fig.~\ref{fig:shearbands}) \cite{yadav2020}.

The force vs. time plot during shear banding is shown in Fig.~\ref{fig:shearbands}(c), with $F_1$ and $F_2$ being the horizontal and vertical forces, respectively. Synchronous high-amplitude oscillations are seen in both the directions, where each oscillation corresponds to formation of a single band. That the oscillations (therefore, shear bands) are highly periodic and equally spaced in time suggests that banding is not a stochastic process. Importantly, the fact that shear bands form in a sequential manner, one at a time, affords opportunities for making time-resolved observations of single band initiation and development.

\begin{figure}
\centering \includegraphics[width=5.8in]{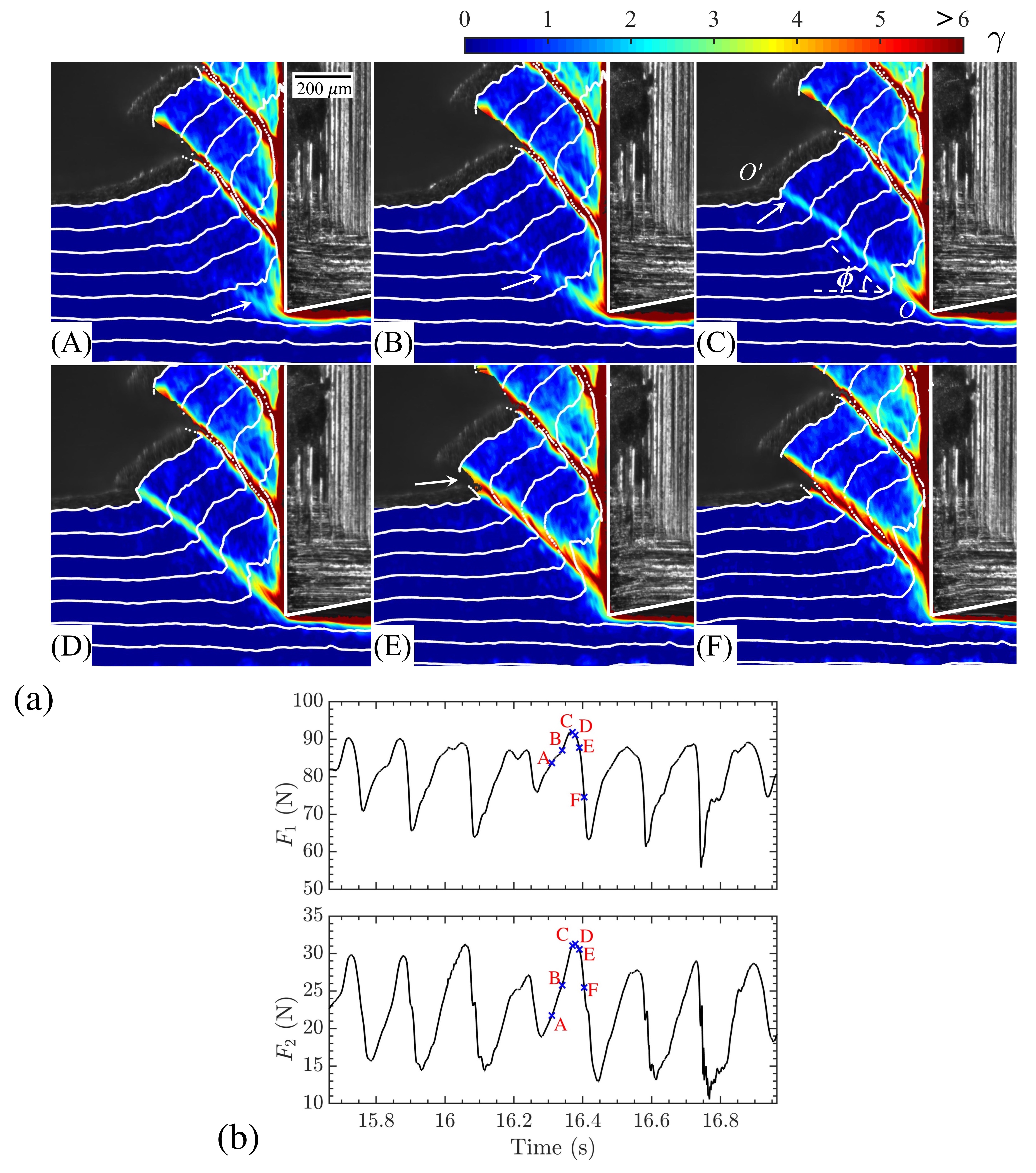}
\caption{\label{fig:dynamics}(a) High-speed image sequence showing formation of a single shear band at $V_0 = 1.5 $ mm/s. Images show the shear strain field distribution along with  streaklines. Frames $A$-$C$ and $D$-$F$ show the shear band initiation and sliding/growth phases, respectively. The time instances corresponding to individual frames are marked on the $F_1$ and $F_2$ force traces in (b). Arrow in frames $A$-$C$ tracks the initiation of shear band, starting from its nucleation at  tool tip (frame $A$) to  propagation towards the material surface and formation of a thin interface $OO'$ (see frame $C$). Formation of this interface $OO'$ coincides with maximum in $F_1$. Large strain accumulation along the interface occurs during the second sliding phase (frames $D$-$F$) under a falling  load. Except for a small crack near the free surface (see at arrow, frame $E$), material remains continuous along the shear band. }
\end{figure}

Figure~\ref{fig:dynamics} shows 6 high-speed frames showing the evolution of a single shear band along with respective time instances on the force plot. Shear band evolves in two distinct stages of initiation and growth. Frames $A$-$C$ show the first stage, involving nucleation of a localized strain inhomogeneity at the tool tip (frame $A$) and propagation of the band front towards the free surface (see at arrow, frame $B$). By frame $C$, a very thin and well-defined band interface $OO'$ traversing the entire specimen is established. At this stage, shear band strain is $\sim 1$ and band orientation is closely aligned with the maximum shear stress direction. For example, the band orientation $\phi$ with respect to $V_0$ is $34^\circ$, while the maximum shear direction calculated based on forces is $36^\circ$. It is also critical to note that the nucleation of band interface (frame $C$)  coincides with the maximum in $F_1$ (see point $C$ in Fig.~\ref{fig:dynamics}(b)).

Frames $D$-$F$ show the second growth stage of shear banding, characterized by accumulation of large strains in the immediate vicinity of band under a dropping load.  As seen from Fig.~\ref{fig:dynamics}, this strain accumulation along the shear band specifically occurs via  relative ``sliding'' of the material blocks on either side of the band plane $OO'$ in equal and opposite directions. 
It is also important to note that, during this sliding process, the material remains continuous across the band despite large strains. It is only during the later stages of sliding that a crack forms at the free surface; for example, see at arrow in frame $E$, where the broken streakline indicates material separation at this location. By frame $F$, it can be seen that large strain levels $> 6$ develop around the shear band, with the band also developing a characteristic thickness of $\sim 20$ $\mu$m. The shear-steps at the free surface are also developed during this sliding stage.

The sequence of steps leading to formation of a fully developed band is now clear. While the nucleation stage establishes band interface and orientation, majority of the localized strains (80\%) develop during the subsequent sliding/growth phase. Once the growth stage comes to a halt, marked by minimum in $F_1$, another band initiates in the neighboring workpiece region, and the process is repeated. 

We also note that the above two-stage mechanism of shear banding is not specific to current model material systems, but has been observed also in other alloys including titanium, cold-worked brass and nickel-base superalloys \cite{sagapuram2016}. In the present study, the two-stage mechanism was also confirmed under different conditions of $V_0$, $\alpha$ and $t_0$. \emph{In situ} observations of banding at $\alpha$ = 20$^{\circ}$ can be found in Fig.~S4 (Supplementary Material), for comparison. 

\subsection{Nucleation stresses} \label{sec:nucleation_stress}
\begin{figure}
\centering \includegraphics[width=4.5in]{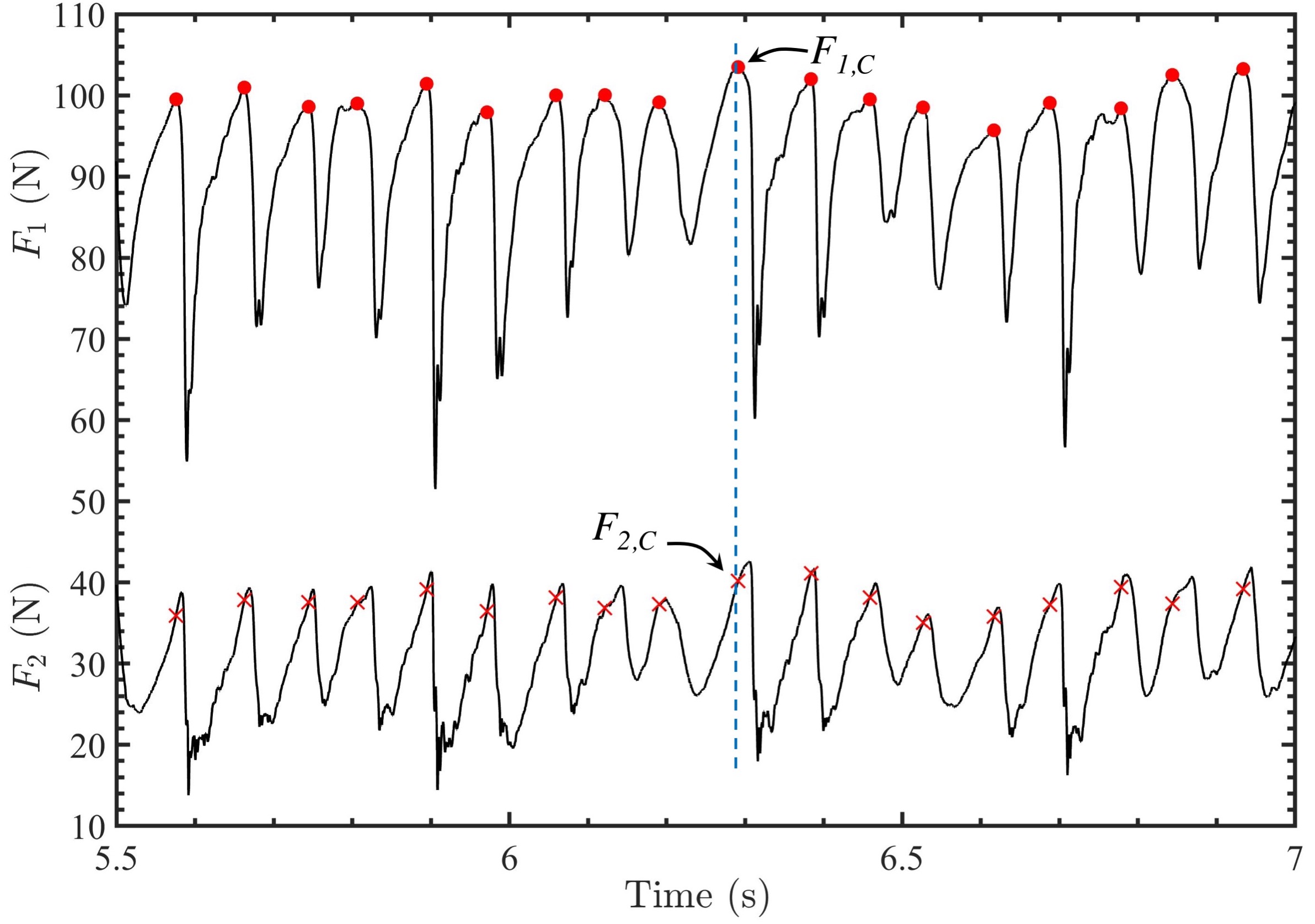}
\caption{\label{fig:peak_force} Typical force trace associated with shear banding ($V_0$ = 4 mm/s,  $\alpha$ = 0$^{\circ}$). Each oscillation in $F_1$ and $F_2$ represents single shear band cycle. $F_{1,C}$ and $F_{2,C}$ represent the forces at band nucleation and are marked using red $\circ$ and $\times$, respectively, in the figure. Average shear band nucleation stress ($\tau_C$) in each experiment was determined from  $F_{1,C}$ and $F_{2,C}$ values collected over multiple bands (see text for details). }
\end{figure}

From the above \emph{in situ} observations, it is clear that shear band nucleation, marking the onset of plastic instability, can be decoupled from the strain-intensive growth stage. Moreover, the fact that nucleation of a thin, well-defined band $OO'$ (see frame $C$, Fig.~\ref{fig:dynamics}) coincides with the peak in the primary force $F_1$  allows us to define and measure the characteristic nucleation stress ($\tau_C$) required for its formation. $\tau_C$ calculation is illustrated using the force plot in Fig.~\ref{fig:peak_force}, where force values at band initiation, $F_{1,C}$ and $F_{2,C}$, are marked using red $\circ$ and $\times$ respectively. It may be noted here, that while band initiation strictly coincides with the maximum in $F_1$, the peak in the tangential force $F_2$ occurs a little after band has initiated. Similar observations were made also in experiments at different $\alpha$ (see Fig.~S4, Supplementary Material). Based on $F_{1,C}$ and $F_{2,C}$, $\tau_C$ for each band can be thus obtained from the resultant shear force parallel to band and  band area. The normal stress $\sigma_C$ acting on the band during its initiation can be similarly derived from the resultant normal force. Mathematically:
\begin{equation*}
{\tau_C} = \dfrac{F_{shear}}{A_S} = \dfrac{(F_{1,C}\cos\phi - {F_{2,C}}\sin\phi)\sin\phi}{bt_0}
\end{equation*}

\begin{equation}
{\sigma_C} = \dfrac{F_{norm}}{A_S} = \dfrac{(F_{1,C}\cos\phi + {F_{2,C}}\sin\phi)\sin\phi}{bt_0}
\label{eq:nucleation_stress}
\end{equation}

\noindent where $F_{shear}$ and $F_{norm}$ are the resolved shear and normal forces respectively, $A_S = bt_0/\sin\phi$ is area of a freshly nucleated band, $b$ is the thickness dimension of the sample, and $\phi$ as before is band orientation with respect to $V_0$ (Fig.~\ref{fig:dynamics}). Measurements over some 200+ bands produced under identical conditions as in Fig.~\ref{fig:dynamics} show that $\tau_C$ is $71 \pm 2$ MPa, while $\sigma_C$ is $91 \pm 3$ MPa. Note that in Eq.~\ref{eq:nucleation_stress}, we do not consider the inertial force components that usually arise in cutting due to  momentum changes across the deformation zone (red shaded area \emph{OA}, Fig.~\ref{fig:cutting})  \cite{shaw1989,albrecht1965}. While this approximation is justified at relatively low speeds used in the present work (since the inertial forces are about 8-10 orders of magnitude lower than the deformation forces, $F_1$ and $F_2$),  one should be careful to include the inertial effects at high cutting speeds  $\gtrsim$ 10 m/s \cite{recht1985}.

\begin{figure}
\centering \includegraphics[width=6.5in]{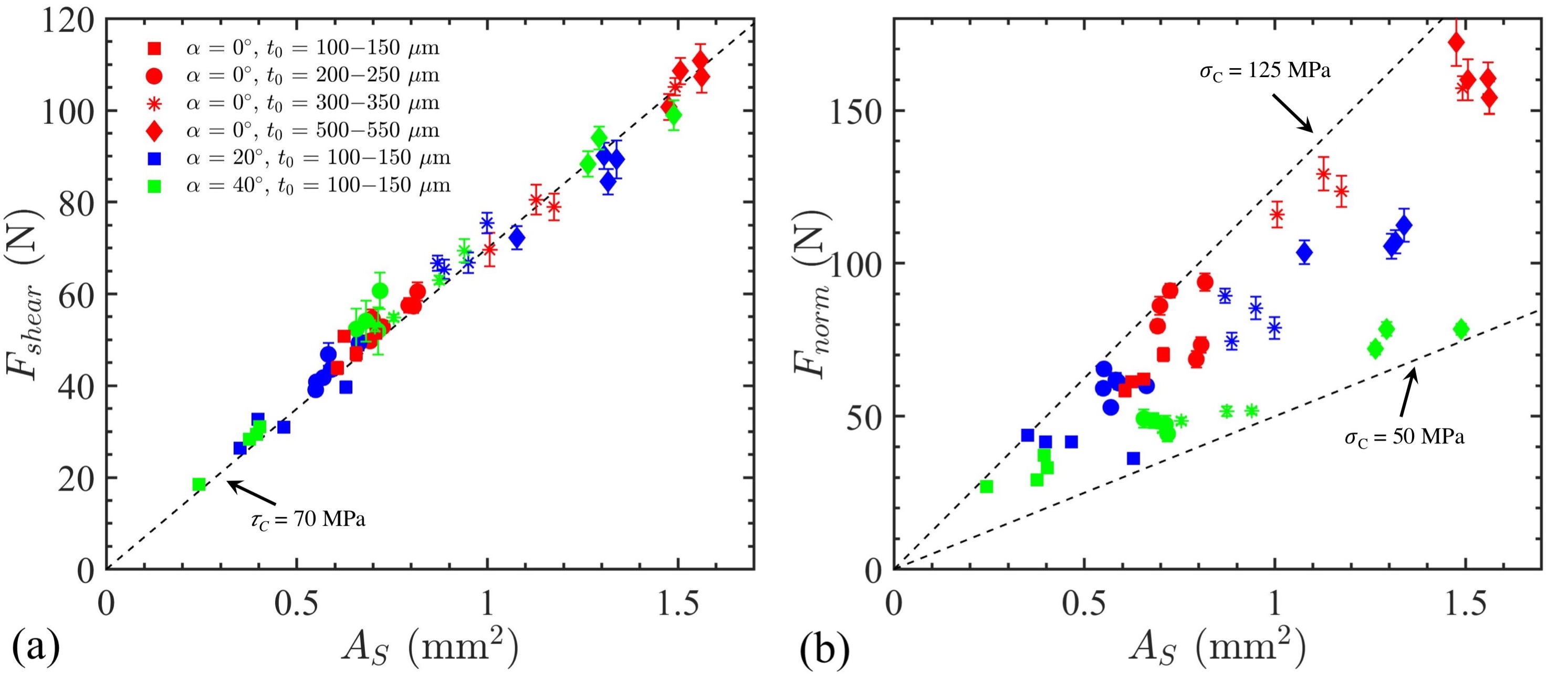}
\caption{\label{fig:universal_plot} Shear band nucleation force data from over 60 different experiments, presented using (a) $F_{shear}$ vs. $A_S$ (b) $F_{norm}$ vs. $A_S$ plots. Each point represents data obtained at a different combination of $\alpha$, $V_0$ and $t_0$. Different colors: red, blue and green represent different $\alpha$: $0^{\circ}$, 20$^{\circ}$ and 40$^{\circ}$ respectively, while different symbols: $\square$, $\circ$, $\ast$ and $\Diamond$ represent different $t_0$ ranges (see figure legend). In (a), all  $F_{shear}$ data fall on a single master line, regardless of the experimental parameters, indicating a constant shear stress at band nucleation. The best straight line fit, shown using the black dashed line, corresponds to $\tau_C = 70$ MPa.  In contrast, $F_{norm}$ data in (b) show a scattered distribution bounded between two straight lines corresponding to $\sigma_C$ of 50 MPa and 125 MPa.}
\end{figure}

Additional experiments over a wide range of testing parameters show that $\tau_C$ is likely a material constant. This is demonstrated in Fig.~\ref{fig:universal_plot}, where the force data ($F_{shear}$ and $F_{norm}$) for band nucleation are plotted as a function of shear band $A_S$, for nearly 60 different experimental combinations of $V_0$, $\alpha$ and $t_0$. Also,  for each experimental condition, force and area values reported are the average of at least 200 bands. From the $F_{shear}$ vs. $A_S$ plot (Fig.~\ref{fig:universal_plot}(a)), it is startling to observe that all the points from various experiments fall on a single master line (black dashed line), whose slope is given by $\tau_C = 70$ MPa. It is also seen that any deviation in the linearity from this line is of the same order of the scatter in the force/area data itself. These observations make clear that $\tau_C$ is an intrinsic material characteristic. 

The dependence of $F_{norm}$ on $A_S$ is shown in Fig.~\ref{fig:universal_plot}(b). It is evident that no single line can be drawn through the data, although it appears that  points align themselves along different families, depending on $\alpha$. In contrast to single $\tau_C$, $\sigma_C$ varies over a 50-125 MPa range. Therefore, it appears that $\tau_C$ also bears no relationship to the normal stress component acting on the band plane. This is clearly akin to the critical resolved shear stress (CRSS) concept  used in crystal plasticity for describing the dislocation slip onset \cite{cottrell1953}.  In fact, additional experiments in the temperature range of -30 $^\circ$C to 60 $^\circ$C showed that  $\tau_C$ remains independent of the ambient temperature and is about 0.05-0.06 times the corresponding shear modulus ($\mu_0$) value, except close the melting point where $\tau_C/\mu_0$ ratio was found to be somewhat higher at 0.08.

It is important to note that the above observations pertaining to  $\tau_C$ are no mere coincidence or peculiar to the idiosyncrasies of Wood's metal, since shear banding in the other two alloys investigated in this study was also characterized by a unique $\tau_C$. Stress data for Alloy 1 ($T_m = 47$ $^\circ$C) and Alloy 3 ($T_m = 138$ $^\circ$C) are shown in Table~\ref{tb:other_materials} at three different $\alpha$ ($V_0 =1$ mm/s).  It is again of interest to note that each material is characterized by a characteristic $\tau_C$ regardless of the normal stress. While the $\tau_C$ values themselves are different depending on the alloy ($\sim 55$ MPa for Alloy 1 and 86 MPa for Alloy 3),   $\tau_C/\mu_0$ ratio was again found to be  $\approx 0.05$ in both the alloys. Although empirical in nature and tested only over a rather limited range of temperatures, these observations suggest a potential scaling relationship between $\tau_C$ and $\mu_0$, similar to that between the theoretical strength  and modulus \cite{kelly1966}.

\subsection{Shear banding to homogeneous flow transition} \label{sec:transition}

\begin{figure}
\centering \includegraphics[width=6.5in]{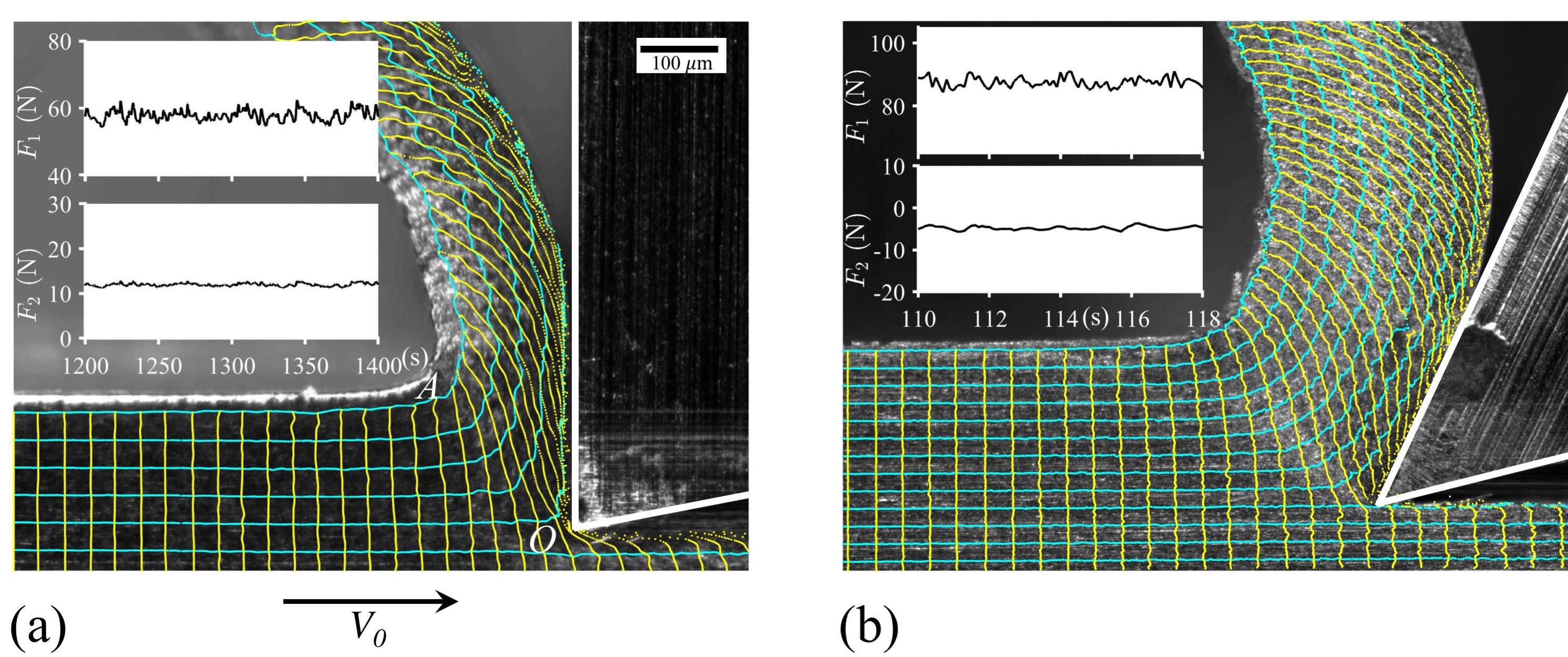}
\caption{\label{fig:transition} Homogeneous flow at lower strain rates of (a) 0.2 /s  and (b) 1 /s. Homogeneous plastic flow devoid of   localization is clearly evident in both the cases from the uniformly deformed grid pattern in the chip and steady-state force profiles (see insets).}
\end{figure}

The observation of constant $\tau_C$ over a range of conditions, coupled with the fact that shear band plane coincides with the maximum shear direction, suggests a condition for banding onset based on  maximum shear stress. This also suggests a transition in the flow mode to homogeneous-type when the shearing stress $\tau$ is $< \tau_C$. This transition was explored in our experiments by altering the material's shear flow stress using external strain rate and temperature as control parameters. As an example, Figure~\ref{fig:transition} shows \emph{in situ} observations of the plastic flow in Wood's metal, where the artificial grid (obtained using PIV flow analysis) is shown superimposed on the raw images. The loading condition in Fig.~\ref{fig:transition}(a) is identical to that in Fig.~\ref{fig:dynamics}, except that $V_0 = 0.01$ mm/s and the corresponding $\dot \gamma \sim 0.1$ /s are two orders lower. Figure~\ref{fig:transition}(b) represents a similar example at $\alpha = 20^\circ $, where $\dot \gamma$ is $\sim 1$ /s, about one order lower than in Fig.~\ref{fig:dynamics}. A striking observation from these experiments is a different flow mode where the undeformed material transforms into a homogeneously deformed chip as the material passes through the plane $OA$. That the underlying flow is ``laminar" without localization can be seen from the smooth streaklines (horizontal cyan lines) and uniform deformation of the grid pattern over the entire chip sample. The PIV strain analysis also confirmed a uniform strain distribution in the chip, with the measured shear strains ($\gamma$) being close to $ 3.5$ and $1.7$ respectively for Figs.~\ref{fig:transition}(a) and (b). Furthermore, this homogeneous flow mode is also accompanied by steady-state forces in both the orthogonal directions ($F_1$ and $F_2$). This is in stark contrast to the shear banding mode, where the force profiles are characterized by high-amplitude, periodic oscillations (Figs.~\ref{fig:shearbands} and \ref{fig:dynamics}).

It is important to note that the principal deformation process in Fig.~\ref{fig:transition} is one of shear deformation confined along a thin zone/plane $OA$. This, for instance, can be clearly observed from the deformation of square grids into elongated rhomboid shapes as material exits the plane $OA$ (Fig.~\ref{fig:transition}). This in turn suggests that the material's shear flow stress $\tau$ along this plane can be estimated using measured forces. In fact, replacing $F_{1,C}$ and $F_{2,C}$ terms in Eq.~\ref{eq:nucleation_stress} with steady-state forces $F_1$ and $F_2$, and $\phi$ with the shear plane orientation with respect to $V_0$, reduces $\tau_C$ calculation to $\tau$ for the steady-state homogeneous flow. This calculation shows that $\tau$ for the two cases shown in Fig.~\ref{fig:transition} are 52 MPa and 58 MPa, respectively. It is evident that these values are well below the critical $\tau_C$ required for band nucleation.

\begin{figure}
\centering \includegraphics[width=4.8in]{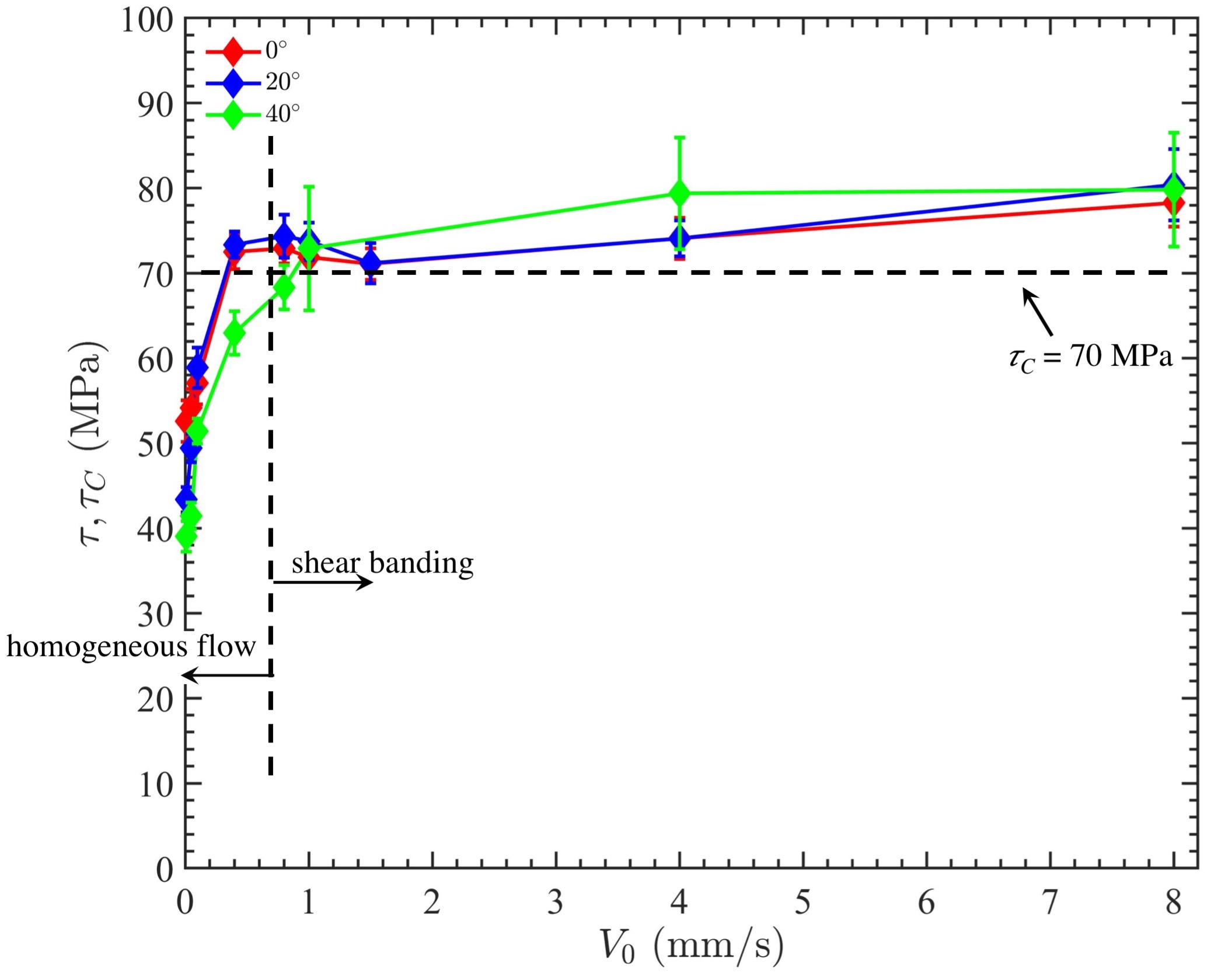}
\caption{\label{fig:shear_stress} Shear stress data shown as a function of $V_0$ for different $\alpha$ (0$^{\circ}$, 20$^{\circ}$ and 40$^{\circ}$).   Approximate boundary demarcating the homogeneous flow and shear banding modes is shown using a vertical black dashed line. Stress values correspond to  shear flow stress ($\tau$) for homogeneous flow and band nucleation stress ($\tau_C$) in the case of shear banding. At low  $V_0$ ($\lesssim 0.5$ mm/s), flow is homogeneous, with  $\tau$ being highly sensitive to $V_0$. Transition to banding occurs when $\tau$ reaches a critical value of about 70 MPa, marked by a horizontal black dashed line. Inside shear banding regime, $\tau_C$ remains independent of $V_0$ or $\alpha$.}
\end{figure}

The transition between the two flow modes and corresponding stress attributes are further elaborated in Fig.~\ref{fig:shear_stress} over a wider experimental dataset, where  shear stresses are plotted as a function of $V_0$, for three different $\alpha$. An approximate boundary between  shear banding and homogeneous flow modes is also marked in the figure using a vertical  dashed line. In the case of homogeneous flow,  stress plotted along the vertical axis represents $\tau$ under which material undergoes steady-state shearing, while for shear banding, it is the critical nucleation stress $\tau_C$.  It is clear from Fig.~\ref{fig:shear_stress} that at small $V_0$, $\tau$ values are significantly lower than $\tau_C$. For example, at 0.01 mm/s, $\tau \sim 40$ MPa, which is well below $\tau_C$. As noted earlier, this is quite consistent with the occurrence of homogeneous flow at small $V_0$. 

From Fig.~\ref{fig:shear_stress}, it can be seen that $\tau$ increases with $V_0$, and when $\tau \simeq \tau_C$ (horizontal dashed line) at $V_0$ of 0.3-1 mm/s, homogeneous flow gives way to shear banding. Within the shear banding domain, $\tau_C$ essentially remains constant at about 70 MPa irrespective of $\alpha$ or $V_0$, consistent with our earlier observations (Fig.~\ref{fig:universal_plot}). Together, these observations suggest a critical shear stress-based condition for the homogeneous flow to shear banding transition.

\begin{figure}
\centering \includegraphics[width=5in]{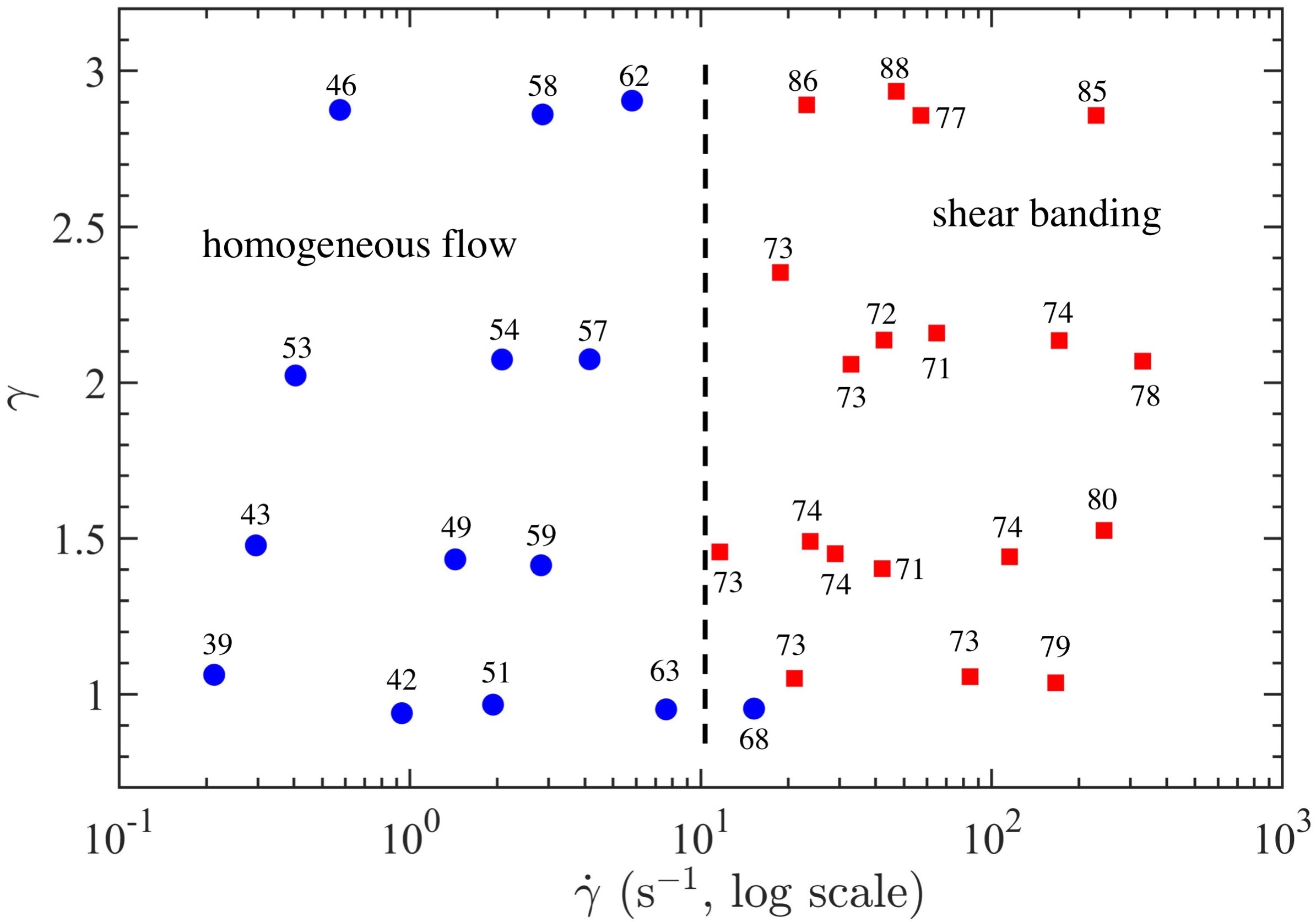}
\caption{\label{fig:transition_map} $\gamma$ vs. $\dot \gamma$ phase diagram showing the domains of homogeneous flow (blue $\circ$) and shear banding (red $\square$) at ambient temperature ($T_0 = 298$ K). Individual points shown correspond to experiments performed at different combinations of $\alpha$ and $V_0$ (correspondingly, $\gamma$ and $\dot \gamma$), while numerical values marked next to points are the corresponding $\tau$ or $\tau_C$ values in MPa. The approximate cut-off $\dot \gamma_C \simeq 10$ /s below which  shear bands do not form is shown by the vertical dashed line. Stress values in the vicinity of this transition are seen to be close to 70 MPa.}
\end{figure}

Based on our experimental observations of the flow type, it is now possible to construct a phase diagram as shown in Fig.~\ref{fig:transition_map}, where the occurrence of each flow mode can be delineated in terms of nominal strain ($\gamma$) and strain rate ($\dot \gamma$) conditions. The scattered points in the diagram are individual experimental runs performed at different combinations of $\alpha$ and $V_0$, corresponding to different combinations of $\gamma$ and $\dot \gamma$. Homogeneous and shear banding modes are marked using blue $\circ$ and red $\square$ symbols respectively, with  corresponding $\tau $ or $\tau_C$ values (in MPa) given next to the symbols. As with the stresses, strain and strain rate data for homogeneous flow correspond to steady-state values, while for shear banding, they represent  pre-localization values just before band nucleation. As seen from the phase diagram, the two flow modes can be clearly demarcated using a vertical dashed line, corresponding to a critical $\dot\gamma_C \sim 10$ /s. In our experiments, simultaneous homogeneous flow and shear band events are sometimes observed close to the cut-off $\dot\gamma_C $, but the region $\dot\gamma < \dot\gamma_C$ shows no shear banding at all, while $\dot\gamma > \dot\gamma_C$ invariably shows shear banding.

It is also of interest to note that in the region $\dot\gamma \simeq \dot\gamma_C$, $\tau \simeq \tau_C$, irrespective of the strain (Fig.~\ref{fig:transition_map}). This should not be surprising given the weak strain-dependence noted for the present alloy (see Fig.~S1). Therefore, the critical stress condition for  flow transition in the present case reduces to a critical rate condition (in the absence of  temperature effects). This equivalence can be also demonstrated  from the alloy's constitutive description ($\tau = K\dot\gamma ^m$ at  298 K), where $ \dot \gamma = \dot \gamma_C = 10$ /s results in a $\tau$ of 75 MPa,  close to the measured $\tau_C$.


\subsection{Temperature effects} \label{sec:temperature}
\begin{figure}
\centering \includegraphics[width=5in]{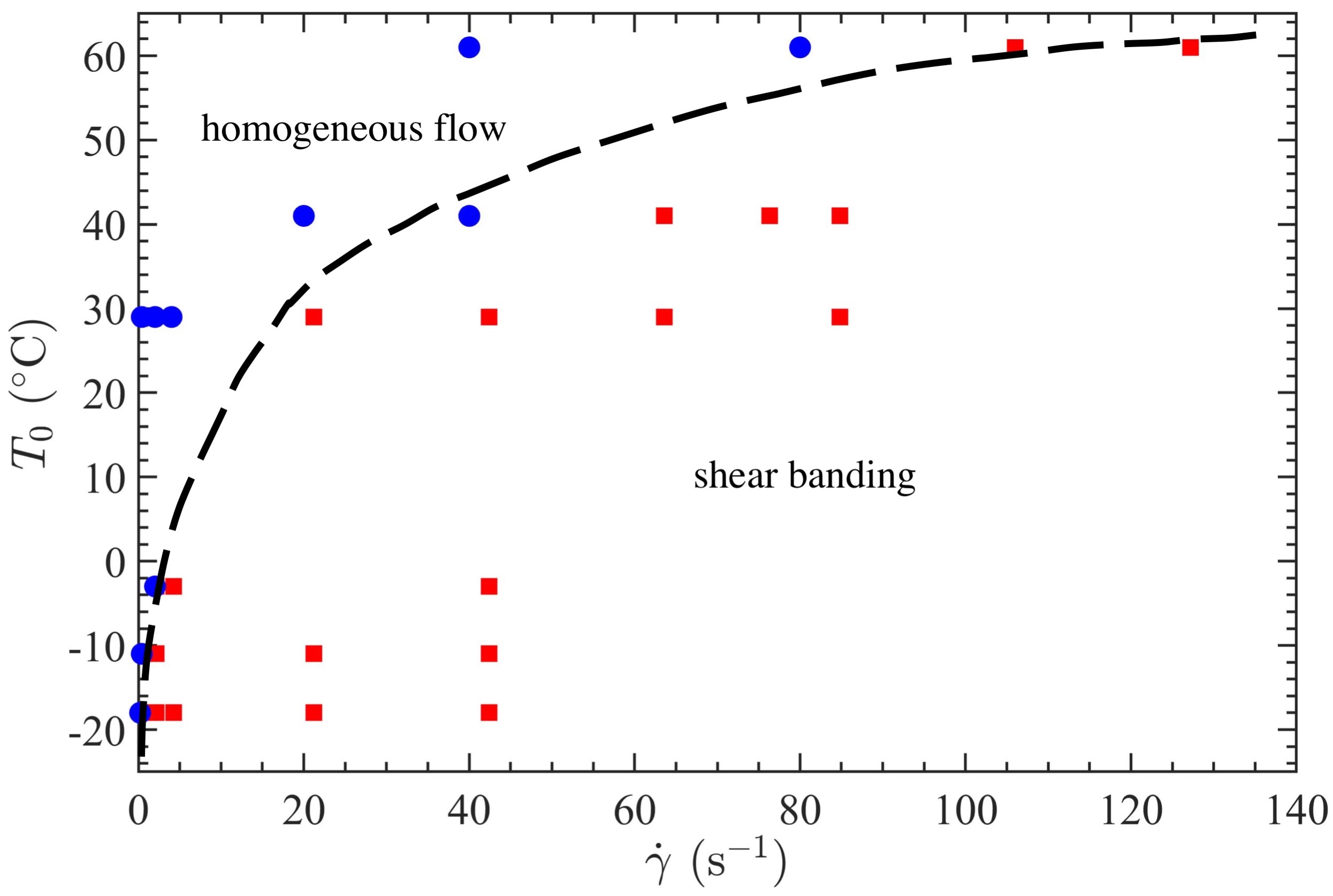}
\caption{\label{fig:temp_transition} $T_0$ vs. $\dot\gamma$ phase diagram showing the occurrence of homogeneous (blue $\circ$) and shear banding (red $\square$) flow modes.  The critical $\dot \gamma_C$ for flow transition is highly sensitive to temperature. This temperature dependence and an approximate boundary separating the flow modes is depicted using the curved black dashed line.}
\end{figure}

The ambient temperature effects on the flow transition and band nucleation characteristics were explored by performing experiments at different temperatures ($T_0$) between -20 $^{\circ}$C and 65 $^{\circ}$C. For Wood's metal, this temperature range corresponds to homologous temperatures ($T_0/T_m$) between 0.74 and 0.98. In these experiments, $\alpha$ was kept constant at 0$^{\circ}$, while $V_0$ was varied between 0.005 and 3 mm/s. Corresponding strain rates are in the 0.1-100 /s range. Results from \emph{in situ} observations of the flow are plotted using $T_0$ vs. $\dot\gamma$ phase diagram, as shown in Fig.~\ref{fig:temp_transition}. As before, the blue $\circ$'s and red $\square$'s represent the homogeneous flow and shear banding modes, respectively. It is clear from the phase diagram that temperature has a remarkable effect on the critical rate $\dot \gamma_C$ at which transition occurs. For example, lowering the temperature to $ -3$ $^\circ$C is seen to reduce $\dot \gamma_C$ to 1 /s, insomuch by an order of magnitude compared to the room-temperature value. In fact, it was found that the critical rate at which flow transition occurs can be approximately described as $\dot \gamma_C \propto \exp(-1/T_0)$, see black dashed line in Fig.~\ref{fig:temp_transition}. We will discuss the origin of this exponential scaling  in Sec.~\ref{sec:discussion}. However, for  present purposes, it is sufficient to note that the measured $\tau$ values in the vicinity of this boundary were all found to be close to $\tau_C$. This suggests that $\tau_C$ is also insensitive to $T_0$, at least in the investigated temperature range.  

Note that in our study, deformation-induced plastic heating and associated temperature rise effects are minimized due to the low cutting speeds (characteristic P\'eclet numbers $\ll 1$). That the temperature rise during cutting is small and no greater than 4-5 $^\circ$C even at the highest $V_0 = 10$ mm/s has been established in our earlier study using thermocouple measurements \cite{yadav2020}. Given that the maximum change in the flow stress ($\sim 2$ MPa) due to this  temperature rise is sufficiently small and well within the experimental uncertainty, the plastic heating effects can be ignored in the present study.

Taken together with the observations in Figs.~\ref{fig:shear_stress} and \ref{fig:transition_map}, this reaffirms that the exact strain, strain rate and temperature conditions that mark the flow transition are all governed by a common $\tau \simeq \tau_C$ condition.

\section{Discussion} \label{sec:discussion}
It is well-established that the condition for intense strain localization along a shear band is that, following the initiation of a thin band plane somewhere in the material, the next increment of strain-hardening is cancelled out by the accompanying material softening, mediated either by deformation-induced temperature rise \cite{recht1964, zener1944} or other microstructural mechanisms \cite{rittel2008, antolovich2014}. The present study has focused on how such a thin plane is initiated in the first place.

Using \emph{in situ} imaging of isolated shear bands, it is shown that shear band initiation occurs by nucleation of strain inhomogeneity at a stress concentration (tool tip in the present case, Fig.~\ref{fig:dynamics}) followed by propagation of its front along the maximum shear direction. The end result is formation of a thin, well-defined plane with a characteristic strain of $\sim 1$ and traversing the whole specimen, see Fig.~\ref{fig:dynamics}. Our force measurements have further shown that there exists a minimum shear stress $\tau_C$ required for nucleating a band, with this stress being a physical characteristic of the material and proportional to the shear modulus. This is clearly brought out in our $\tau_C$ measurements over a wide range of experimental conditions (Fig.~\ref{fig:universal_plot}), and also demonstrated across three alloys with different melting points (Table~\ref{tb:other_materials}). That at stress levels $ < \tau_C$, a different flow mode --- homogeneous flow --- without shear bands occurs is also unambiguously established using our \emph{in situ} experiments (Fig.~\ref{fig:transition}). The  domains where  homogeneous and shear banding flow modes  operate are illustrated using phase diagrams (Figs.~\ref{fig:transition_map} and \ref{fig:temp_transition}). 

\begin{figure}
\centering \includegraphics[width=5.0in]{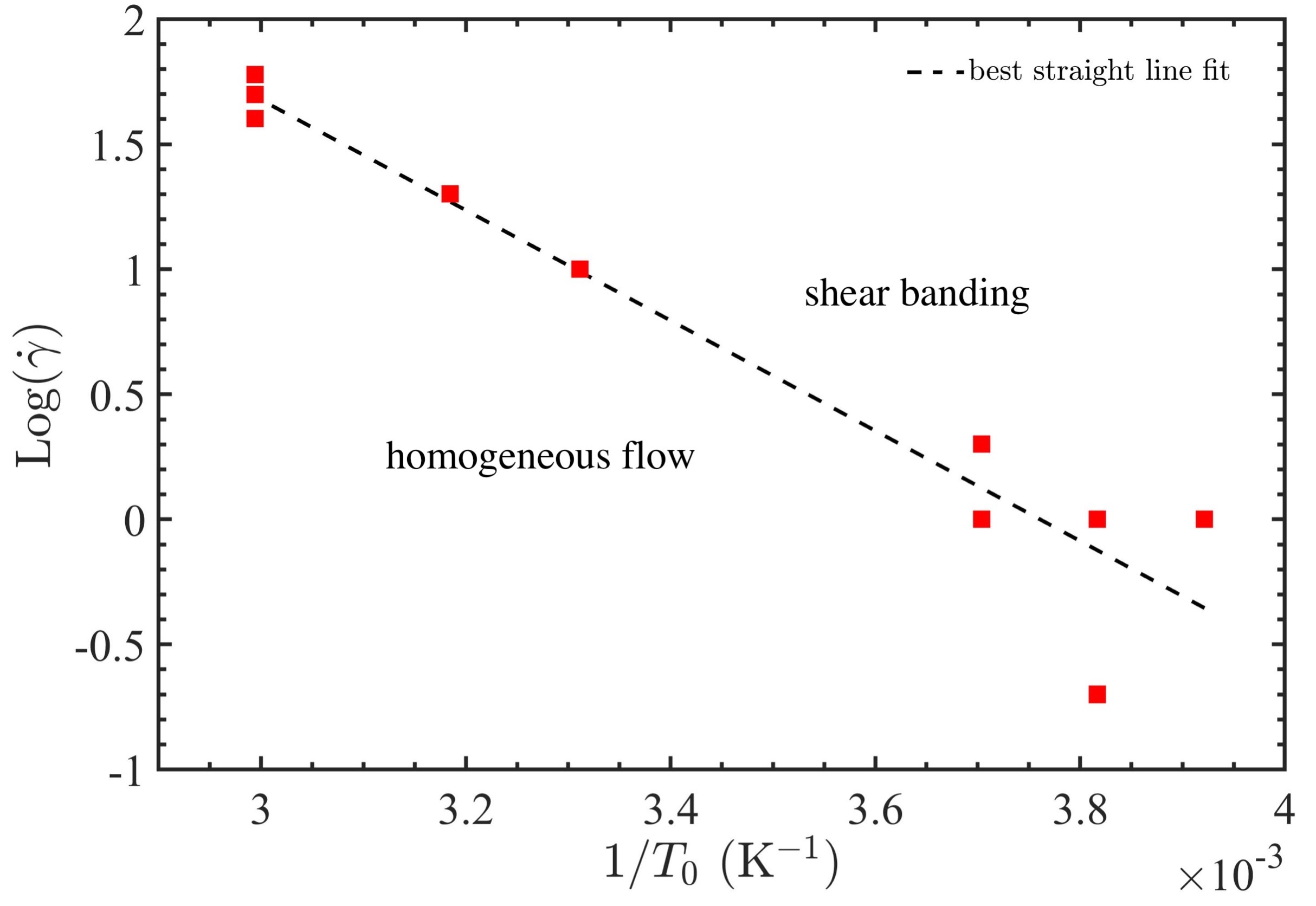}
\caption{\label{fig:scaling} Log$(\dot\gamma)$ vs. $1/T_0$ plot of the points close to the  transition boundary in Fig.~\ref{fig:temp_transition}. The points are seen to closely fall on a single straight line, indicating that the critical $\dot \gamma_c$ for flow transition scales as $\exp(-1/T_0)$. The best straight line fit to the data is shown as a black dashed line.}
\end{figure}

The critical stress condition for the flow transition was demonstrated in Figs.~\ref{fig:shear_stress} and \ref{fig:transition_map} using direct stress measurements. Additional support for this condition from a physical standpoint is presented in Fig.~\ref{fig:scaling}, where the critical rate-temperature combinations marking the flow transition are plotted as $\log (\dot\gamma)$ vs. $1/T_0$. It is seen that all the data fall on a  straight line, with the deviations being within experimental uncertainty of accurately determining the critical $\dot\gamma_C$ for flow transition.  In fact, it turns out that a generalized thermo-viscoplastic material with a flow stress dependence of the form $\tau \propto \left\{ \dot \gamma \exp(Q/RT)\right\}^m$ \cite{zener1944} exhibits exactly this type of linear scaling between  rate and temperature at a constant $\tau$. In this picture, $Q$ is the activation energy for plastic flow, $R$ is the universal gas constant, and $m$ as before is the rate sensitivity. For the data plotted in Fig.~\ref{fig:scaling}, the best straight line fit (black dashed line) is obtained at $\tau = 73$ MPa and $Q = 42$ kJ/mol. Clearly, this $\tau$ is close to the $\tau_C$ measured independently in our other experiments (Fig.~\ref{fig:universal_plot}). Equally importantly, $Q$ is close to the activation energy for self-diffusion for all the constitutive elements \cite{belchuk1990}. This is suggestive of an underlying plastic flow mechanism where the rate-controlling step is vacancy diffusion to or from the climbing dislocations held up at discrete obstacles \cite{kocks1975}.

Based on  foregoing observations and analysis, a plausible microscopic mechanism by which a thin shear band plane nucleates can be now discussed. Consider, for instance, dislocation-mediated plastic flow connected with dislocation jumps across potential obstacles (e.g., precipitates, second-phase particles, forest dislocations), and let $\tau_C$ be the shear stress required to set the dislocation in motion around the obstacles in the absence of any thermal contribution. At temperatures above absolute zero, even if the applied stress is less than $\tau_C$, dislocation can overcome the obstacle with the help of thermal fluctuations, provided sufficient time is made available for it to make the jump. However, at high strain rates, the specimen's internal plastic rate can lag the external displacement rate, and when stress reaches a critical value, one can envision breakdown of piled-up dislocations along a plane as a sudden `burst'. In this picture, shear band nucleation at a critical $\tau_C$ is associated with strain transients resulting from sudden breakdown of dislocation barriers. If the temperature rise or structural changes accompanying the pile-up breakthrough are such that they overcome the strain/strain-rate hardening effect, strain localization results along a freshly nucleated band. While shear band initiation based on this type of dislocation pile-up `avalanche' mechanism has been postulated before by several authors \cite{gilman1994,armstrong1989,hirth1992,coffey1989} based on theoretical considerations, we believe our study provides the first quantitative experimental support in terms of  activation energy and nucleation stress. It is of interest to note that formation of microscopic slip bands is also known to be mediated by similar dislocation pile-up breakdown mechanisms \cite{nabarro1967}.  

It is noteworthy that the above microscopic description of shear band nucleation is consistent with our earlier observations that normal stresses play a little role in band nucleation, as expected for any dislocation-mediated process. Furthermore, the fact that $\tau_C$ scales with $\mu_0$ is  in line with the classical rate-dependent plasticity theory where the athermal stress component is often taken to be proportional to the shear modulus \cite{campbell1970, kocks1975}. Importantly, this description also provides a unified explanation for several important observations related to shear banding in metals. For example, factors that promote shear banding such as multi-phase or precipitation-hardened microstructures (as opposed to single-phase) \cite{ duffy1992}, very low temperatures \cite{ basinski1957}, high deformation rates  and high levels of cold-work \cite{ hatherly1984} ---  all of which restrict dislocation motion --- can be understood from the above dislocation basis. However, the precise mechanism of how dislocation slip that is confined to specific crystallographic planes can result in sample-scale bands that cut across several grains remains an open problem \cite{ brown2005}.

It is pertinent to also briefly discuss possible generalization of our results to other metals, beyond the model alloys investigated in this study. In this context, we compare our results with prior, albeit limited, data on band nucleation stresses. Using dynamic torsional experiments, Duffy's group reported $\tau_C$ values for steels in the range of 350-1100 MPa, depending on the material's microstructure, composition and prior heat treatment \cite{ duffy1992}. The corresponding $\tau_C/\mu_0$ ratios are between 0.01-0.02, substantially lower than observed in the present study. It appears therefore that, as with regular dislocation nucleation \cite{borovikov2017}, the activation energy, and therefore, characteristic stresses required for band nucleation depend considerably on the type and nature of the obstacles/barriers themselves. This hypothesis clearly requires additional study. However, our current work has established the correlation between band dynamics and the corresponding force trace (Fig.~\ref{fig:dynamics}) so that $\tau_C$ measurements can be made across different material systems by merely observing the force traces, without need for detailed \emph{in situ} imaging. Such measurements would be of practical value as they can be used in conjunction with the activation energies to predict the critical strain, strain rate and temperature conditions for the  localized flow onset.

Lastly, the dislocations Burgers vector and the fact that dislocations do not `run away' after breaking through the obstacles but are characterized by a finite (stress-dependent) velocity \cite{gilman1994} introduce intrinsic length- and time-scales into the band nucleation problem. In this regard, detailed examination of shear displacements and band propagation velocities  during shear band nucleation, and their possible connection with dislocation dynamics at the microscale, will be likely valuable for placing shear banding within a wider dislocation physics framework.

\section{Conclusions} \label{sec:conclusions}
The nucleation of isolated shear bands in metals has been studied using a special shear loading configuration and \emph{in situ} imaging using high-speed photography. This enabled  band nucleation  to be decoupled from the subsequent strain-intensive growth phase. Based on synchronous imaging and force measurements, the existence of a critical shear stress ($\tau_C$) required for band nucleation was established across three different alloy systems. The independence of $\tau_C$ with regards to normal stresses, its constancy over a wide range of testing conditions, and scaling with the shear modulus ($\mu_0$): $\tau_C \approx 0.05 \mu_0$, were also demonstrated. It is shown that these results can be rationalized from the viewpoint that shear band nucleation is a mechanical instability of the crystal lattice arising due to sudden breakdown of dislocation barriers. A smooth, homogeneous flow mode without shear bands was reproduced at stress levels $< \tau_C$. Based on these observations, phase diagrams that delineate the strain rate, strain and temperature domains under which each of the flow modes (homogeneous vs. shear banding) occur were developed. It is shown that the observed boundaries demarcating the two flow modes are consistent with  $\tau \simeq \tau_C$ condition.

\section*{Acknowledgments}
We would like to thank Harshit Chawla of Texas A\&M University for his assistance with image processing and PIV analysis. We would also like to acknowledge support from US DOE  award DE-NA0003525 (via Sandia National Laboratories sub-award 2016312).


%
{\bibliography{bibFile}}



 \begin{sidewaystable}
\begin{center}
 \begin{tabular}{lccc} 

 \hline \hline
\textbf{Properties} & Alloy 1 & Alloy 2 & Alloy 3 \\
 \hline \hline
\textbf{Melting Point ($T_m$)} & 47 $^\circ$C & 70 $^\circ$C & 138 $^\circ$C\\
\textbf{Composition (wt. \%)} & 44.7Bi, 22.6Pb, 19.1In, 8.3Sn,  5.3Cd &  50Bi, 26.7Pb, 13.3Sn,  10Cd  & 58Bi, 42Sn \\

\textbf{Density ($\rho$)} &  $8.98 \times 10^3 $ kg/m$^3$ & $9.26 \times 10^3 $ kg/m$^3$  & $8.15 \times 10^3 $ kg/m$^3$  \\
\textbf{Thermal Diffusivity ($\kappa$)} & 0.09 $\times $10$^{-4}$ m$^2$/s & 0.14 $\times $10$^{-4}$ m$^2$/s & 0.14 $\times $10$^{-4}$ m$^2$/s  \\
\textbf{Specific Heat ($C$)} & 180.6 W$\cdot$s/kg$\cdot$K & 165.7 W$\cdot$s/kg$\cdot$K & 167.7 W$\cdot$s/kg$\cdot$K \\
\textbf{Thermal Conductivity ($k$)} & 14.4 W/m$\cdot$K & 21.5 W/m$\cdot$K & 19.4 W/m$\cdot$K  \\
\textbf{Strain-rate sensitivity ($m$)} &0.12 & 0.13 & 0.06  \\
 \hline \hline
\end{tabular}
\end{center}
\caption{Thermophysical properties of low melting-point alloys at 23$^\circ$C.} 
\label{tb:properties}
\end{sidewaystable}

\begin{sidewaystable}
\begin{center}
 \begin{tabular}{lccc} 

 \hline \hline
{} & $\alpha = 0^\circ$ & $\alpha = 20^\circ$ & $\alpha = 40^\circ$ \\
 \hline \hline

\textbf{Alloy 1 } & $\sigma_C$ = 105 $\pm$ 4 MPa, $\tau_C$ = 52 $\pm$ 3 MPa & $\sigma_C$ = 79 $\pm$ 2 MPa, $\tau_C$ = 56 $\pm$ 2 MPa & $\sigma_C$ = 68 $\pm$ 3 MPa, $\tau_C$ = 57 $\pm$ 2 MPa  \\

\textbf{Alloy 2 } & $\sigma_C$ = 115 $\pm$ 3 MPa, $\tau_C$ = 72 $\pm$ 2 MPa & $\sigma_C$ = 103 $\pm$ 4 MPa, $\tau_C$ = 74 $\pm$ 2 MPa & $\sigma_C$ = 62 $\pm$ 4 MPa, $\tau_C$ = 73 $\pm$ 7 MPa \\

\textbf{Alloy 3 } & $\sigma_C$ = 222 $\pm$ 8 MPa, $\tau_C$ = 92 $\pm$ 4 MPa & $\sigma_C$ = 136 $\pm$ 6 MPa, $\tau_C$ = 86 $\pm$ 2 MPa & $\sigma_C$ = 95 $\pm$ 7 MPa, $\tau_C$ = 86 $\pm$ 4 MPa \\

 \hline \hline
\end{tabular}
\end{center}
\caption{ Nucleation stress data for the three  alloys at different $\alpha$. Values reported in the table are the average and one standard deviation calculated from multiple shear bands. $V_0$ = 1 mm/s.} 
\label{tb:other_materials}
\end{sidewaystable}



\end{document}


\begin{center}
\noindent \textbf{Supplementary Material: Nucleation properties of isolated shear bands}

\noindent Authors: Shwetabh Yadav and Dinakar Sagapuram

\noindent Texas A\&M University, College Station, TX 77843, USA
\end{center}

\section{Mechanical behavior of low melting point alloys}

Mechanical properties of the low temperature melting point alloys used in this study were studied by carrying compression tests (ASTM E9-09) on cylindrical shaped samples (10 mm dia., 20  mm height). Petroleum jelly was used as a lubricant between the sample and the compression platen to minimize friction and barreling. The strain-rate effects on the flow behavior were studied at room temperature by  varying the compression speed between 0.001 mm/s to 1 mm/s; the corresponding nominal strain rates ($\dot\varepsilon$) were in the range of 10$^{-5}$ /s to 10$^{-2}$ /s. The effect of ambient testing temperature was also studied for one of the alloys (Alloy 2) by carrying out similar tests at different temperatures ($T_0$)  of -30 to 60 $^\circ$C at a constant  strain rate of 2.5 $\times$ 10$^{-3}$ /s. For these latter experiments, a special testing setup was used where both the compression platens and sample were mounted within a furnace to maintain   constant ambient temperature throughout the test. Temperatures below the room temperature were obtained by regulating the flow of liquid nitrogen inside the furnace chamber.

\begin{figure}[H]
\centering \includegraphics[width=6.5in]{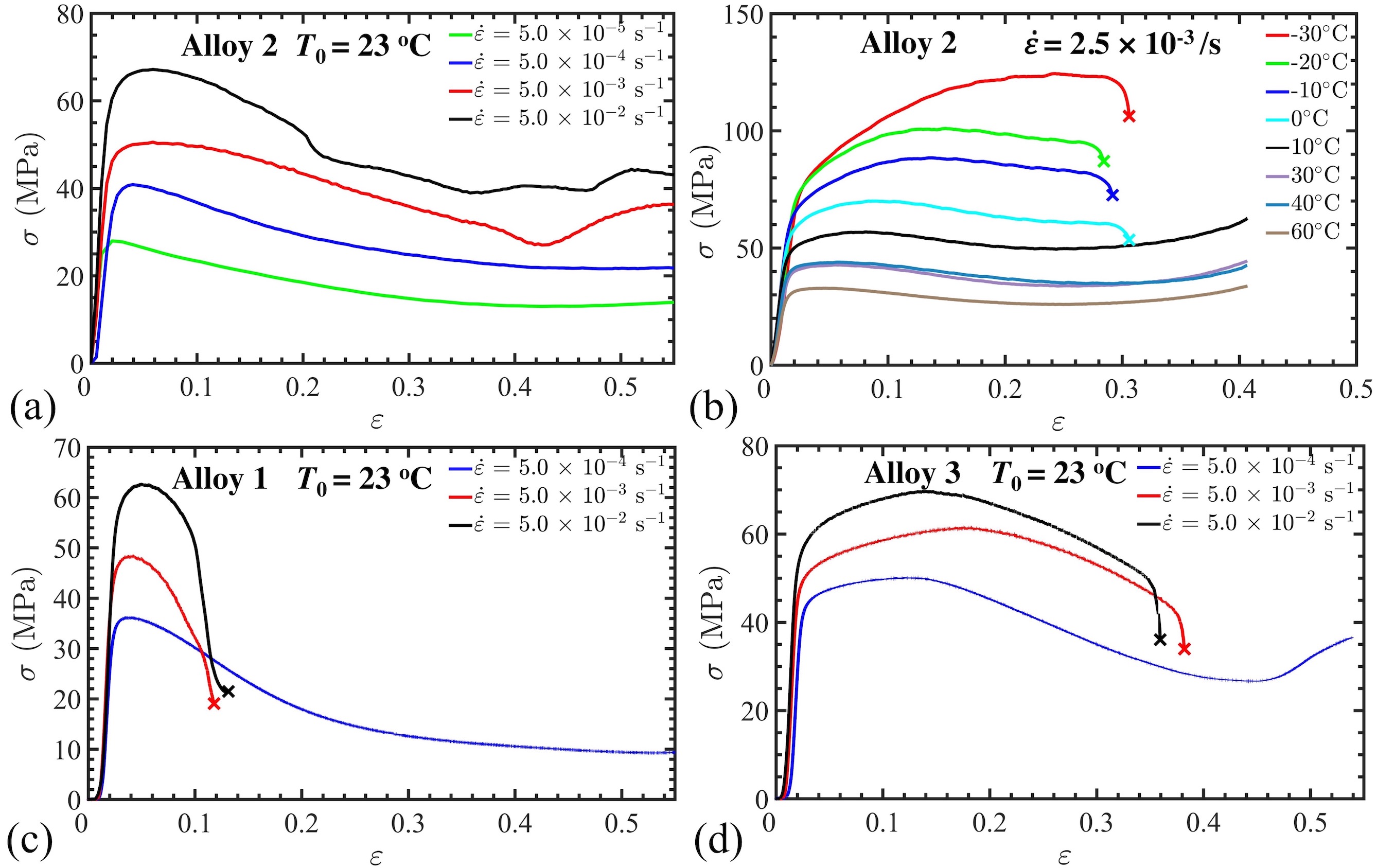}
\caption{\label{fig:compression}Compression stress-strain curves  for Alloy 2 at (a) different strain rates (at a constant $T_0 = 23$ $^\circ$C) and (b) different ambient temperatures (at a constant nominal $\dot\varepsilon$ of 2.5 $\times$ 10$^{-3}$ /s). (c) and (d) show the corresponding data for Alloy 1 and Alloy 3 at room temperature. The points marked with $\times$ denote sample fracture.}
\end{figure}

Figures~\ref{fig:compression}(a) and (b) show the (true) stress-strain plots for Alloy 2 ($T_m = 70$ $^\circ$C) at different strain rates and temperatures, respectively. The alloy shows a strong dependence on both the strain rate and temperature. At a constant temperature  ($T_0 = 23$ $^\circ$ C), the  material exhibits flow-softening behavior post-yielding, with the flow stress being a strong function of the strain rate, see Fig.~\ref{fig:compression}(a). For a given strain, it is found that the flow stress varies with strain rate as a power law:  $\sigma \propto \dot\varepsilon^m$, with $m \sim$ 0.13. With respect to the temperature (Fig.~\ref{fig:compression}(b)), decreasing $T_0$ at a constant strain rate is found to have a hardening effect,   both in terms of the yield stress as well as the stress dependence on strain.  The temperature-dependence of yield stress ($\sigma_y$) is found to be roughly of the form:  $\sigma_y \propto \exp(A/T_0)$, where $A$ is a  constant of about $ 938$ K. 

Similar tests performed at a constant $T_0 = 23$ $^\circ$C  for Alloy 1 and 3 also show a strong rate-dependence (Fig.~\ref{fig:compression}(c) and (d)), with the corresponding $m$ values being $ 0.12$ and 0.06, respectively. Alloy 1 with a lower melting point ($T_m = 47$ $^\circ$C) exhibits post-yielding softening behavior similar to Alloy 2,  while  Alloy 3 ($T_m = 138$ $^\circ$C) is characterized by a maximum in the stress-strain curve at a strain of $\sim 0.15$.

\section{Shear band formation at $\alpha$ = 20$^{\circ}$}

\begin{figure}[H]
\centering \includegraphics[width=5.5in]{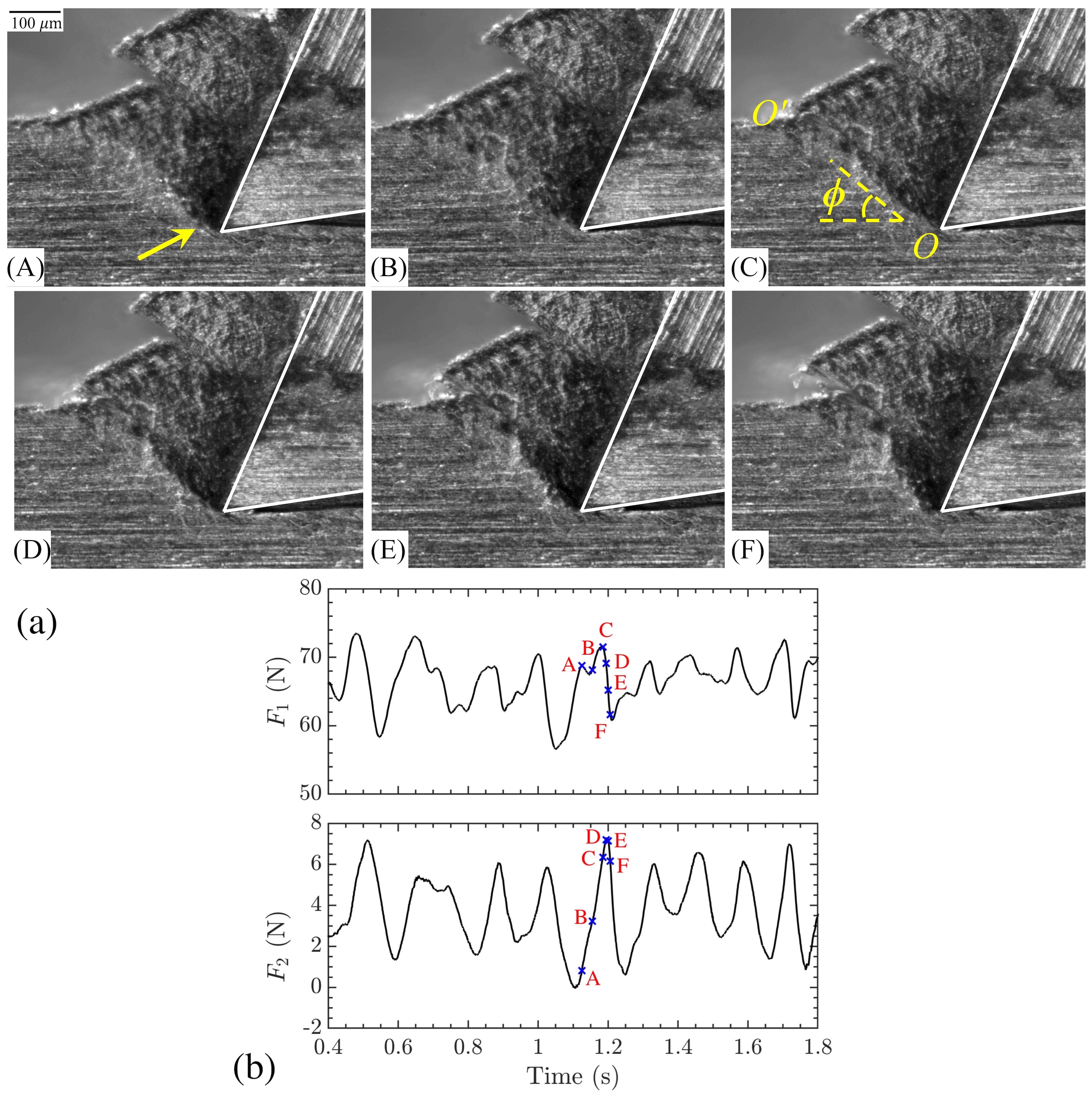}
\caption{\label{fig:dynamics}Shear band dynamics in Wood's metal (Alloy 2) at $\alpha = 20^\circ$. (a) High-speed image sequence showing two-stage mechanism of shear banding.  Frames $A$-$C$ and $D$-$F$ show the shear band initiation and sliding phases, respectively.  (b) The corresponding force profile, with time instances of frames $A$-$F$ marked on the $F_1$ and $F_2$ plots.}
\end{figure}

The two-stage mechanism of shear band formation shown in Fig.~3 of the main manuscript is common across different experimental parameters.  Figure~\ref{fig:dynamics} shows single shear band formation in Wood's metal (Alloy 2) at $\alpha = 20^\circ$, where 6 high-speed image frames and the corresponding time instances on the force trace are shown. All the essential features of banding dynamics --- band nucleation  at the tool tip (arrow, frames $A$), propagation towards the free surface to establish a well-defined band interface (frame $C$), and sliding along this interface (frames $D$-$F$) --- are replicated in this experiment. It is also of interest to note that the formation of interface $OO'$ coincides with the peak in $F_1$, and sliding occurs under a dropping load. Nucleation stress calculations for the present case shows $\tau_C = 77$ MPa, consistent with the $\tau_C$ values from other experiments (see Fig.~5, main manuscript). As before, the band orientation $\phi$ ($46^\circ$) is again closely aligned with the maximum shear direction, which is at $43^\circ$ with respect to $V_0$ direction.

